\documentclass{pasj00}
\usepackage{times}

\begin{document}
\SetRunningHead{Totani et al.}{Damping Wing Analysis of GRB 130606A}
\Received{2013/12/14}%{yyyy/mm/dd}
\Accepted{2014/2/6}%{yyyy/mm/dd}
%\Published{}%{yyyy/mm/dd}

\title{Probing Intergalactic Neutral Hydrogen by 
  the Lyman Alpha Red Damping Wing of Gamma-Ray Burst 130606A
  Afterglow Spectrum at $z = 5.913$}

\author{%
   Tomonori \textsc{Totani},\altaffilmark{1,2}
   Kentaro \textsc{Aoki},\altaffilmark{3}
   Takashi \textsc{Hattori},\altaffilmark{3}
   George \textsc{Kosugi},\altaffilmark{4}
   Yuu \textsc{Niino},\altaffilmark{5}
   Tetsuya \textsc{Hashimoto},\altaffilmark{5}
   Nobuyuki \textsc{Kawai},\altaffilmark{6}
   Kouji \textsc{Ohta},\altaffilmark{7}
   Takanori \textsc{Sakamoto},\altaffilmark{8}
   and
   Toru \textsc{Yamada}\altaffilmark{9}}
   \altaffiltext{1}{Department of Astronomy, The University of Tokyo, 
     Hongo, Tokyo 113-0033}
   \altaffiltext{2}{Research Center for the Early Universe, The University of 
     Tokyo, Hongo, Tokyo 113-0033}
   \altaffiltext{3}{Subaru Telescope, National Astronomical Observatory of 
   Japan, 650 North A'ohoku Place, Hilo, HI 96720, USA}
   \altaffiltext{4}{Chile Observatory, National Astronomical Observatory
    of Japan, Mitaka, Tokyo 181-8588}
   \altaffiltext{5}{Optical and Infrared Astronomy Division, National 
    Astronomical Observatory, Mitaka, Tokyo 181-8588}
   \altaffiltext{6}{Department of Physics, Tokyo Institute of Technology, 
    2-12-1 Ookayama, Meguro-ku, Tokyo 152-8551} 
   \altaffiltext{7}{Department of Astronomy, Kyoto University, Sakyo-ku, 
    Kyoto 606-8502} 
   \altaffiltext{8}{Department of Physics and Mathematics, College of 
    Science and Engineering, Aoyama Gakuin University, 5-10-1 Fuchinobe, 
    Chuo-ku, Sagamihara-shi Kanagawa 252-5258} 
   \altaffiltext{9}{Astronomical Institute, Tohoku University,
    Aoba-ku, Sendai 980-8578} 
%% `\KeyWords{}' always has to be placed before `\maketitle'.
%Do NOT move this preamble from here!

\maketitle

\begin{abstract}
The unprecedentedly bright optical afterglow of GRB 130606A located by
{\it Swift} at a redshift close to the reionization era ($z = 5.913$)
provides a new opportunity to probe the ionization status of
intergalactic medium (IGM). Here we present an analysis of the red
Ly$\alpha$ damping wing of the afterglow spectrum taken by
Subaru/FOCAS during 10.4--13.2 hr after the burst.  We find that the
minimal model including only the baseline power-law and
H\emissiontype{I} absorption in the host galaxy does not give a good
fit, leaving residuals showing concave curvature in 8400--8900 {\AA}
with an amplitude of about 0.6\% of the flux.  Such a curvature in the
short wavelength range cannot be explained either by extinction at the
host with standard extinction curves, intrinsic curvature of afterglow
spectra, or by the known systematic uncertainties in the observed
spectrum. The red damping wing by intervening H\emissiontype{I} gas
outside the host can reduce the residual by about 3$\sigma$
statistical significance.  We find that a damped Ly$\alpha$ system is
not favored as the origin of this intervening H\emissiontype{I}
absorption, from the observed Ly$\beta$ and metal absorption
features. Therefore absorption by diffuse IGM remains as a plausible
explanation.  A fit by a simple uniform IGM model requires
H\emissiontype{I} neutral fraction of $f_{\rm H\emissiontype{I}} \sim$
0.1--0.5 depending on the distance to the GRB host, implying high
$f_{\rm H\emissiontype{I}}$ IGM associated with the observed dark
Gunn-Peterson (GP) troughs. This gives a new evidence that the
reionization is not yet complete at $z \sim 6$. 
\end{abstract}

\section{Introduction}

Hydrogen is the dominant component of the baryonic matter in the
universe and most of it is in the form of diffuse intergalactic medium
(IGM).  Observations of the cosmic microwave background radiation
proves that it became neutral around 400,000 years after the Big Bang,
but the absorption strength by IGM neutral hydrogen in quasar spectra
[i.e., the Gunn-Peterson (GP, Gunn \& Peterson 1965) test] tells us
that IGM in the present universe is highly ionized. It is widely
believed that the cosmic reionization occurred around $z \sim 6$--10
by radiation emitted from stars of the earliest generations, but the
detailed history of reionization and early galaxy formation is yet to
be observationally revealed.  Detecting neutral hydrogen before
reionization is an important challenge to understand the first stage
of galaxy formation (see, e.g., Fan et al. 2006; Barkana \& Loeb 2007;
Robertson et al. 2010; Fan 2012 for reviews).

Although IGM hydrogen seems largely ionized at $z \lesssim$ 5.5 from
the quasar GP test, the GP optical depth rapidly increases with
redshift around $z \sim$ 6 indicating a systematic change of the IGM
nature that may be related to reionization (Fan 2012).  However,
completely saturated absorptions (GP troughs) give only a weak lower
bound on the neutral fraction of IGM hydrogen, $f_{\rm
  H\emissiontype{I}} \equiv n_{\rm H\emissiontype{I}} / n_{\rm H} 
\gtrsim 10^{-4}$.  Furthermore, quasars are strongly biased tracers
generally found in the strongest density peaks in large-scale
structure, and quasars may have altered the ionization status of
surrounding IGM by their strong radiation. Therefore, there is
currently no direct evidence that reionization was complete by $z
\sim$ 5--6 (Mesinger 2010).

It is thus important to search neutral IGM having higher $f_{\rm
  H\emissiontype{I}}$ than that probed by the GP test at $z \sim 6$,
and it may be detected by the damping wing feature redward of the
resonant Ly$\alpha$ wavelength. Gamma-ray burst (GRB) afterglows are
expected to be a unique and useful tool for this approach
(Miralda-Escude 1998; Lamb \& Reichart 2000; Zhang 2007) because of
the following reasons.  The intrinsic nonthermal synchrotron spectrum
before absorption is a simple power-law, and hence a precise analysis
on the damping wing shape is possible. GRBs are expected to be less
biased than quasars, allowing to probe more normal regions in large
scale structure.  Finally, their radiation hardly affects the
ionization status of surrounding IGM because of much shorter duration
than quasars.

However, strong constraints on reionization from GRBs have not yet
been obtained, because of the low event rate of sufficiently bright
GRBs at high redshifts, and contamination of hydrogen in their host
galaxies to the damping wing feature.  The only GRB-based constraint
on reionization is a weak upper limit of $f_{\rm H\emissiontype{I}} <
0.6$ at $z = 6.3$ (Kawai et al. 2006; Totani et al. 2006), where the
observed red Ly$\alpha$ damping wing can be explained only by neutral
hydrogen in its host galaxy with a column density of $N_{\rm
  H\emissiontype{I}} \sim 4 \times 10^{21} \ \rm cm^{-2}$.  GRBs at
even higher redshifts have also been detected (Greiner et al. 2009;
Patel et al. 2010; Salvaterra et al. 2009; Tanvir et al. 2009;
Cucchiara et al. 2011), but their spectra do not have a sufficient
signal-to-noise ratio for a precise damping wing analysis.

Recently a new opportunity has been given by GRB 130606A, which was
detected by {\it Swift} and KONUS-{\it Wind} (Ukwatta et al. 2013;
Golenetskii et al. 2013).  The redshift was determined to be $z =
5.913$ (Castro-Tirado et al. 2013a; Chornock et al. 2013) by its
exceptionally bright afterglow among high redshift GRBs, and
furthermore, a low H\emissiontype{I} column density in the host
($N_{\rm H\emissiontype{I}} \sim 7 \times 10^{19} \ \rm cm^{-2}$,
Chornock et al. 2013; Castro-Tirado et al. 2013b) implies that a much
better constraint on IGM neutral hydrogen may be obtained than the
previous cases.  Here we present a damping wing analysis of the
unprecedentedly high signal-to-noise optical spectrum of GRB 130606A
taken by the Subaru Telescope during 10.4--13.2 hr after the burst, to
derive new constraints on the neutral hydrogen fraction of IGM at $z
\sim 6$.

We describe our observation and data reduction in Section
\ref{section:obs}, and the damping wing analysis is presented in
Section \ref{section:DW}, showing an evidence for intervening
H\emissiontype{I} absorption in addition to that in the host galaxy.
Various sources of systematic uncertainties are examined in Section
\ref{section:systematics}, and we examine the possibility of an
intervening damped Ly$\alpha$ system (DLA) as the origin of the
intervening H\emissiontype{I} gas in Section \ref{section:DLA}. The
summary and conclusions are presented in Section
\ref{section:conclusions} with some discussions.  Throughout the
paper, H\emissiontype{I} column density $N_{\rm H\emissiontype{I}}$ is
given in units of cm$^{-2}$.

\section{Observation and Spectrum Reduction}
\label{section:obs}

The afterglow of GRB~130606A was observed on 2013 June 7 (UT) with the
optical spectrograph FOCAS (Kashikawa et al. 2002) attached to the
Subaru 8.2-m telescope (Iye et al. 2004). After the observation of GRB
050904 afterglow (Kawai et al. 2006) by the same instrument, FOCAS was
upgraded with a new red-sensitive CCD in June 2010, resulting in a
$\sim$30\% higher quantum efficiency at 9000 \AA \ and significantly
reduced fringing.  We obtained three types of spectra using the 300R,
VPH650, and VPH900 grisms, covering 5820--10345, 6055--7747, and
7472--10538 \AA \ with the pixel scales of 1.35, 0.61, and 0.74 \AA,
respectively.  The 0.8 arcsec slit width mask was used for all the
spectra, and the O58 filter was used for 300R and VPH900, while Y47
for VPH650.  The list of each 20-minutes integration during 9.3--16.5
hr after the burst is summarized in Table \ref{table:obs}.  The
atmospheric dispersion corrector was not available at this time, and
we set the slit position angle at 90 degree (i.e., east-west
direction) so that it is close to the parallactic angle when the
airmass is large.  It was a photometric condition with good seeing
(0.5--0.7 arcsec FWHM in optical), which is smaller than the slit
width.  Consequently, the spectral resolution is 6.9 and 3.9~{\AA}
FWHM for the 300R and VPH900/650 spectra, respectively.  The data were
reduced using IRAF\footnote {IRAF is distributed by the National
  Optical Astronomy Observatory, which is operated by the Association
  of Universities for Research in Astronomy (AURA), Inc., under
  cooperative agreement with the National Science Foundation.  } for
the procedures of overscan and bias subtraction, flat-fielding,
wavelength calibration, and sky subtraction.  Wavelength calibration
was done by using OH night sky emission lines, with the resulting rms
calibration error of 0.1--0.2 \AA. The spectrum was then converted to
the vacuum wavelength.

\begin{figure}
\begin{center}
\includegraphics[width=110mm,angle=-90]{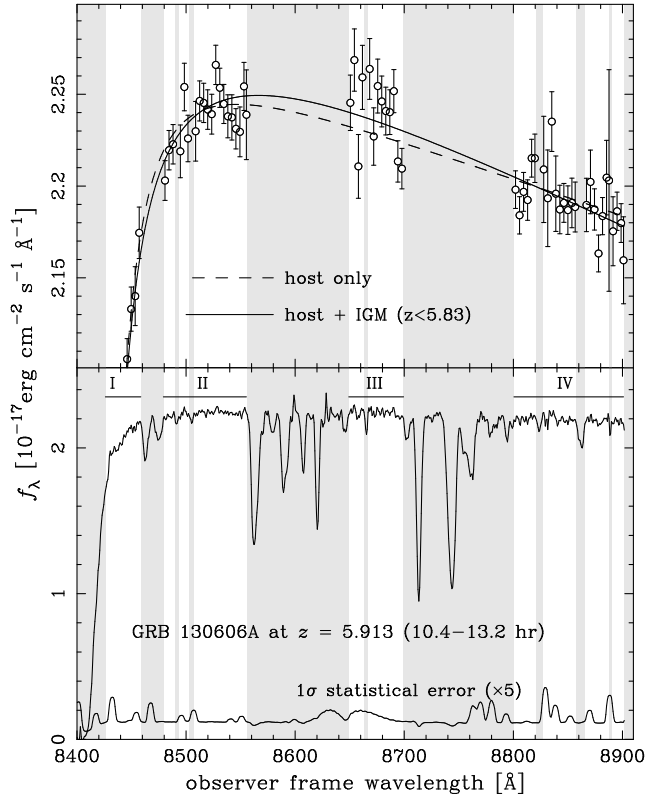}
\end{center}
\caption{The optical spectrum of GRB 130606A afterglow, around the
  redshifted Ly$\alpha$ and its red damping wing region.  In the
  bottom panel, the observed spectrum per pixel and its error are
  shown, where the error is multiplied by a factor of 5 for the
  presentation purpose.  In the top panel, the best-fit curves of the
  host-only and IGM-low$z_u$ models are compared with the spectral
  data binned up to five pixels, which were used in the $\chi^2$
  calculation.  The wavelength regions excluded from the damping wing
  analysis are indicated by the shaded regions, and the defined
  wavelength ranges I--IV are indicated in the lower panel.  }
\label{fig:spec_obs}
\end{figure}

We use the VPH900 spectrum taken during 10.4--13.2 hr after the burst
for the main damping wing analysis reported in this work
(Fig. \ref{fig:spec_obs}).  The fractional statistical errors of the
spectrum are $\sim$1\% per pixel (0.74 \AA), allowing us to examine
the spectral shapes on 100 {\AA} scale with a 0.1\% level
statistical precision.  We paid a special attention to the data
reduction as follows, since the shape of the damping wing may be
affected even by small systematic uncertainties in such a high
precision test.  In this work we are interested in the narrow
wavelength range including the damping wing, and we performed the
reduction limited within the range of 8377--8902 \AA, to avoid
unnecessary systematic bias induced by, e.g., a sensitivity curve
fitting in a wider wavelength range.  Wavelength longer than 8902 \AA
\ was not used because of stronger atmospheric emission/absorption.
As for the flux calibration, our primary data is calibrated using the
spectral standard star Feige 34 in the CALSPEC database established as
the Hubble Space Telescope (HST) calibration system. The Subaru data
of Feige 34 obtained in the same night are compared with the CALSPEC
template, and the sensitivity curve is constructed by fitting with a
quadratic polynomial.  The fit is good and we do not find any residual
beyond statistical errors. Therefore, a 0.1\% level precision test is
possible for 100\AA-scale spectral shapes within the reduced
wavelength range.

\begin{table}
\small
  \caption{The Subaru/FOCAS Observation Log}
\label{table:obs}
\begin{center} 
\begin{tabular}{cccc}
\hline
\hline
\# & mode & June 7 (UT)$^*$ & Hours after GRB trigger$^*$\\
\hline
 1 & 300R & 06:25:59.7 &  9.35574 \\
 2 & 300R & 06:46:30.6 &  9.69767 \\
 3 & 300R & 07:07:01.1 & 10.03946 \\
\hline
 4 & VPH900 & 07:28:00.2 & 10.38921 \\
 5 & VPH900 & 07:48:31.8 & 10.73134 \\
 6 & VPH900 & 08:09:00.7 & 11.07269 \\
 7 & VPH900 & 08:32:27.9 & 11.46359 \\
 8 & VPH900 & 08:52:59.0 & 11.80554 \\
 9 & VPH900 & 09:13:29.0 & 12.14722 \\
10 & VPH900 & 09:36:54.7 & 12.53770 \\
11 & VPH900 & 09:57:25.7 & 12.87964 \\
12 & VPH900 & 10:17:55.9 & 13.22135 \\
\hline
13 & VPH650 & 10:47:10.8 & 13.70882 \\
14 & VPH650 & 11:07:40.6 & 14.05044 \\
15 & VPH650 & 11:28:11.2 & 14.39228 \\
16 & VPH650 & 11:52:04.5 & 14.79042 \\
17 & VPH650 & 12:12:36.0 & 15.13249 \\
18 & VPH650 & 12:33:06.3 & 15.47425 \\
19 & VPH650 & 12:56:58.8 & 15.87216 \\
20 & VPH650 & 13:17:28.5 & 16.21374 \\
21 & VPH650 & 13:37:58.7 & 16.55546 \\
\hline
\hline
\end{tabular}
\end{center}
$^*$ Midpoint of 20-min exposures.
\end{table}

\section{Damping Wing Fitting Analysis}
\label{section:DW}

\subsection{Model Description}
\label{section:model}

The observed spectrum is fitted by models including absorptions by
neutral hydrogen along the sightline, imprinted on the baseline
power-law with an index $\beta$ ($f_\nu \propto \nu^\beta$).  We
consider three components of H\emissiontype{I} absorption:
H\emissiontype{I} in the host galaxy, diffuse H\emissiontype{I} in
IGM, and an intervening DLA that is often found in high-redshift
quasar/GRB spectra.

The host H\emissiontype{I} component is assumed to have a single
Gaussian radial velocity dispersion $\sigma_v$ with a column density
$N_{\rm H\emissiontype{I}}^{\rm host}$.  In our baseline analysis, we
fixed the redshift of host H\emissiontype{I} \ as $z_{\rm GRB} =
5.9131$, from the three low-ionization metal lines found in the same
VPH900 spectrum used for the damping wing fit.  (We found
Si\emissiontype{II} $\lambda$1260.42, O\emissiontype{I}
$\lambda$1302.17, and C\emissiontype{II} $\lambda$1334.53 and their
redshifts are $z = 5.91305 \pm 0.00002$, $5.91307 \pm 0.00005$, and
$5.91270 \pm 0.00002$, respectively. The C\emissiontype{II} redshift
may be affected by a slightly blending nearby line.)  The widths of
these lines are $\sim$5 \AA \ FWHM that is slightly extended beyond
the spectral resolution (3.9 \AA), indicating a velocity dispersion of
$\sim$100 km/s FWHM (or 43 km/s in $\sigma$).  Therefore we perform
the fit in the limited range of $\sigma_v = $ 0--100 km/s, though the
fit results are almost insensitive to $\sigma_v$ (see next section).

We consider a component of constant-density diffuse IGM with the
damping wing formula of Miralda-Escude (1998) with a neutral fraction
$f_{\rm H\emissiontype{I}}$ extending in a redshift range from $z_{\rm
  IGM,l}$ to $z_{\rm IGM,u}$.  We adopted the GP optical depth
$\tau_{\rm GP} = 3.97 \times 10^5 f_{\rm H\emissiontype{I}}
[(1+z)/7]^{3/2}$ calculated from the latest estimates of cosmological
parameters (Komatsu et al. 2011) and the primordial helium abundance
(Peimbert et al. 2007).  Though the distribution of neutral hydrogen
in IGM should be clumpy and inhomogeneous in reality, we test this
simple model to get a rough estimate for mean $f_{\rm
  H\emissiontype{I}}$ required to explain the observed damping wing.
Dark Ly$\alpha$ troughs are found at $z \gtrsim 5.67$ along the
sightline to this GRB (Chornock et al. 2013), and hence we set the
lower bound of the H\emissiontype{I} distribution as $z_{\rm IGM,l} =
5.67$, but our results do not significantly change by choosing
different values, because the contribution to the damping wing from
H\emissiontype{I} \ around $z_{\rm IGM, l}$ is small. For the upper
bound, we will test two models; one is assuming $z_{\rm IGM,u} =
z_{\rm GRB}$, while the other treats $z_{\rm IGM,u}$ as a free
parameter.

If we find an evidence for absorption by intervening H\emissiontype{I}
gas outside the host galaxy, it may be a result of an intervening DLA
rather than diffuse IGM.  Low-ionization metal absorption lines are
generally associated to DLAs, and this would give a constraint on the
DLA scenario. For the GRB 130606A spectrum, a metal absorption system
at $z = 5.806$ is found by Si\emissiontype{II}, O\emissiontype{I}, and
C\emissiontype{II} lines (Chornock et al. 2013), which is a good
candidate of the DLA redshift.  We will thus perform a fit to the
observed damping wing with the DLA component at $z_{\rm DLA} = 5.806$,
with a column density of $N_{\rm H\emissiontype{I}}^{\rm DLA}$.

Therefore fits are performed with the free model parameters of
$\beta$, $N_{\rm H\emissiontype{I}}^{\rm host}$, $\sigma_v$, $f_{\rm
  H\emissiontype{I}}$, $z_{\rm IGM,u}$ and $N_{\rm
  H\emissiontype{I}}^{\rm DLA}$, while the overall normalization
factor of the model is always chosen to fit the data (i.e.,
marginalized). The theoretical fluxes are calculated for each data
point (i.e., pixel) of the VPH900 spectrum, then corrected for the
Galactic extinction ($A_{V}=0.066$ mag, Schlafly \& Finkbeiner 2011),
and finally convolved with a Gaussian profile of the instrumental
wavelength resolution (3.9 \AA \ FWHM).

\begin{figure}
\begin{center}
\includegraphics[width=77mm,angle=-90]{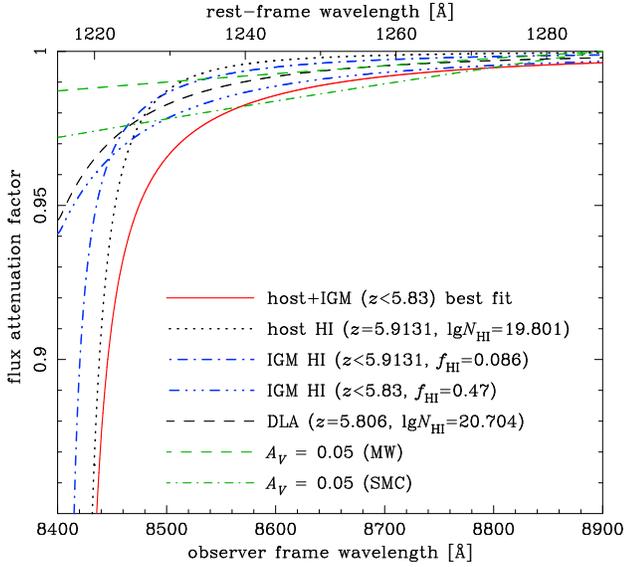}
\end{center}
\caption{Flux attenuation factor of the Ly$\alpha$ damping wing by
  various H\emissiontype{I} components and extinction at the host.
  The red solid curve is the total attenuation by H\emissiontype{I} in
  IGM and the host galaxy, for the IGM-low$z_u$ model in Table
  \ref{table:parameters}.  The IGM and host components of this model
  are also separately shown.  The IGM absorption in the IGM-$z_{\rm
    GRB}$ model and the DLA component of the intervening DLA model are
  also shown for comparison.  The green curves are the effect of
  introducing extinction by dust in the host galaxy (normalized to
  unity at 8900 \AA), using the MW or SMC extinction curves
  (Fitzpatrick 1999; Gordon et al. 2003). The value of $A_V = 0.05$
  was chosen for the presentation purpose; the amount of dust expected
  from $N_{\rm H\emissiontype{I}}$ and metallicity of the host galaxy
  is much smaller ($A_V \lesssim 0.01$, see Section
  \ref{section:model}).  }
\label{fig:spec_model}
\end{figure}

We do not consider possible extinction by dust in the host, because
standard extinction curves have a monotonic wavelength dependence
within the short range analyzed here, and adding nonzero extinction is
almost perfectly absorbed by an according change of the index $\beta$,
giving a practically identical fit (see Fig. \ref{fig:spec_model}
where the absorption profiles of various components are presented).
Therefore the index $\beta$ here is not the intrinsic value for the
afterglow emission, but should be regarded as an effective value in
the analyzed wavelength range including extinction at the host.
However, it is also useful to estimate the amount of extinction
expected from the hydrogen column density in the host, $N_{\rm
  H\emissiontype{I}} \sim 7 \times 10^{19} \ \rm cm^{-2}$.  Using the
mean relations for the Milky Way (MW, Predehl \& Schmitt 1995) and the
Small Magellanic Cloud (SMC, Gordon et al. 2003), this column density
translates into $A_V = 0.036$ and 0.0048, respectively.  The value
from the MW relation is likely an overestimate for this GRB, because
the optical spectra of this GRB indicate a metallicity lower than 0.3
or 1/7 times the solar abundance (Chornock et al. 2013; Castro-Tirado
et al. 2013b). It should also be noted that many GRB afterglows show
smaller extinction than those expected from soft X-ray absorption or
the $N_{\rm H\emissiontype{I}}$-$A_V$ relations (Schady et al. 2010),
implying that the above estimate of $A_V$ is likely an overestimate.
For reference, $\Delta \beta$ = $-$0.042 and $-$0.094 are required to
represent the reddening of $A_V = 0.01$ in 8400--8900 \AA \ by the
change of $\beta$, for the MW and SMC extinction curves, respectively.

\begin{figure}
\begin{center}
\includegraphics[width=90mm]{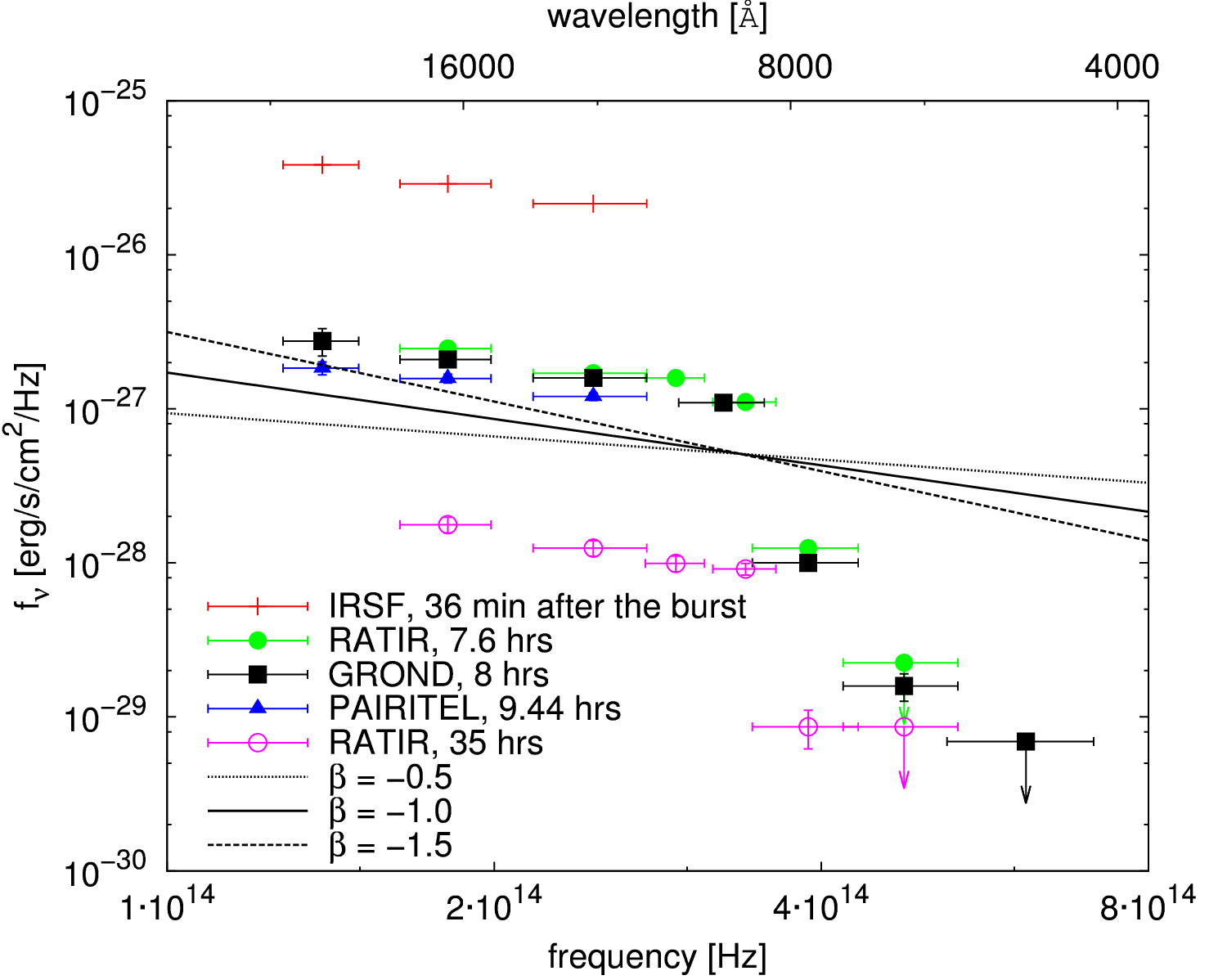}
\end{center}
\caption{Reported optical/near-infrared photometric measurements for
  the GRB 130606A afterglow, by observations of IRSF ($JHK_s$,
  Nagayama et al. 2013), RATIR ($r'i'ZYJH$, Butler et al. 2013a,b),
  PAIRITEL ($JHK_s$, Morgan et al. 2013), and GROND ($g'r'i'z'JHK$,
  Afonso et al. 2013). Power-laws with indices $\beta = -0.5, -1$ and
  $-1.5$ are shown, where the flux is normalized to that observed by
  our Subaru spectrum ($f_\lambda = 1.9 \times 10^{-17} \ \rm erg
  \ cm^{-2} s^{-1}$ \AA$^{-1}$ at 8900 \AA \ during 10.4--13.2 hr
  after the burst).  }
\label{fig:NIR_SED}
\end{figure}

Other independent observational constraints are available for the
power-law index $\beta$ from near-infrared photometric observations of
the afterglow.  In Fig. \ref{fig:NIR_SED} we plot the reported
optical/near-infrared photometric observations (Nagayama et al. 2013;
Butler et al. 2013a,b; Morgan et al. 2013; Afonso et al. 2013),
showing a constant value of $\beta \sim -1$ at wavelengths longer than
the Lyman break from 36 min to 35 hr after the burst.  Using
observations at the time close to the Subaru observation, we find
$\beta = -0.94 \pm 0.05$ from the GROND $JHK$ band data (8 hrs) and
$-0.78 \pm 0.10$ from the PAIRITEL $JHK_s$ band data (9.4 hrs).  As
mentioned above, extinction at the host should be at most $A_V \sim
0.01$, and the expected change of $\beta$ between the near-infrared
bands and our wavelength range should be $\Delta \beta \lesssim 0.05$.
The best-fit $\beta$ values obtained by our fits presented below (see
Table \ref{table:parameters}) are roughly consistent with these
independent measurements, taking into account the systematic
uncertainties about host extinction and in estimating $\beta$ from
photometric measurements (e.g., band filters, response function, or
zero-point calibration).

\subsection{Fitting Procedures and Statistical Significance Estimations}

The spectrum has been reduced in the range of 8377--8902 \AA, but we
further removed the wavelength shorter than 8426 \AA \ from the
fitting analysis, because this region is dominated by the host
H\emissiontype{I} \ absorption rather than the other interesting
components (Fig. \ref{fig:spec_model}).  This is the region where the
damping wing rapidly drops with decreasing wavelength, and a fitting
becomes sensitive to the velocity distribution of H\emissiontype{I}
atoms in the host.  Although we included a Gaussian distribution with
the $\sigma_v$ parameter in our fit, the realistic distribution is
unlikely a pure Gaussian (Wolfe et al. 2005).  The fit in this range
is also sensitive to the uncertainties about the mean redshift of
H\emissiontype{I} in the host (i.e., $z_{\rm GRB}$) and the
instrumental spectral resolution. The lower bound of 8246 {\AA} was
determined as the shortest wavelength with which the fitting results
are mostly insensitive to $\sigma_v$, as can be seen in Table
\ref{table:parameters}.  Discernible absorption features were also
removed from the fitting as shown in Fig \ref{fig:spec_obs}.  Then 323
pixels remain for the fitting analysis.

The statistical errors are estimated from the noise spectrum produced
by the object+sky electron counts and readout noise, which is random
without correlation among different pixels.  (Note that the FWHM of
spectral resolution is larger than one pixel, but the random
statistical errors appear in each pixel in addition to the signal
convolved by the spectral resolution.)  However, pixel-by-pixel
statistical fluctuation of the final spectrum is smaller than the
noise estimate, because the fluxes at neighboring pixels are mixed to
some extent by the resampling for wavelength calibration in the
spectrum reduction.  At the same time, correlation between the errors
of neighbouring pixels is generated, which may affect a $\chi^2$
analysis.  This problem can be solved by binning the five connecting
pixels, and we confirmed that the fit residuals of the best-fit model
(the IGM-low$z_u$ model, see below) do not show any correlation
between neighbouring bins within the statistical uncertainties.  There
are 68 bins after the binning, and we can adopt the standard $\chi^2$
analysis to calculate the best-fit values and 
the statistical confidence regions of the model parameters.

\subsection{Fitting Results}

\begin{table*}
\small
%\scriptsize
\caption{The Best Fit Parameters of the Model Fittings$^*$}
\label{table:parameters}
\begin{center}
\begin{tabular}{lccccc}
\hline
\hline
model  & $\beta$ & $\lg (N_{\rm H\emissiontype{I}}^{\rm host})^\dagger$ &
 $\sigma_v$ (km/s)$^\ddagger$ & $f_{\rm H\emissiontype{I}}$ 
   or $\lg (N_{\rm H\emissiontype{I}}^{\rm DLA})^\dagger$ 
 & $\chi^2$ \\
\hline
host H\emissiontype{I} only  
  & $-1.12^{+0.05}_{-0.04}$ &  19.866$^{+0.009}_{-0.010}$ & 0.0$^{+63.1}_{-0.0}$ 
  & - & 87.58 \\
+IGM ($z \le z_{\rm GRB}$)  
  & $-0.93^{+0.04}_{-0.04}$ &  19.719$^{+0.040}_{-0.040}$ 
  & 100.0$^{+0.0}_{-48.4}$ & 0.086$^{+0.012}_{-0.011}$ & 78.66 \\
+IGM ($z \le 5.83$)  
  & $-0.74^{+0.09}_{-0.07}$ &  19.801$^{+0.023}_{-0.023}$ 
  & 100.0$^{+0.0}_{-70.4}$  & 0.47$^{+0.08}_{-0.07}$ & 76.62 \\
+DLA ($z = 5.806$)  
  & $-0.81^{+0.10}_{-0.04}$ &  19.799$^{+0.026}_{-0.026}$ 
  & 100.0$^{+0.0}_{-58.4}$ & 20.704$^{+0.066}_{-0.067}$ & 77.02 \\
\hline
\hline
\end{tabular}
\end{center}
$^*$The fit results for the host-only, IGM-$z_{\rm GRB}$,
IGM-low$z_u$, and intervening DLA models.  The quoted errors
are statistical 1$\sigma$. There are 68 data points used in the $\chi^2$
calculation. \\
$^\dagger$The neutral hydrogen column density in the GRB host
($N_{\rm H\emissiontype{I}}^{\rm host}$) and that in an intervening
DLA ($N_{\rm H\emissiontype{I}}^{\rm DLA}$) are in units of cm$^{-2}$. \\
$^\ddagger$The survey range of $\sigma_v$ is limited
to 0--100 km/s, motivated from the observed widths of the metal
absorption lines.

\end{table*}

\begin{figure}
\begin{center}
\includegraphics[width=95mm,angle=-90]{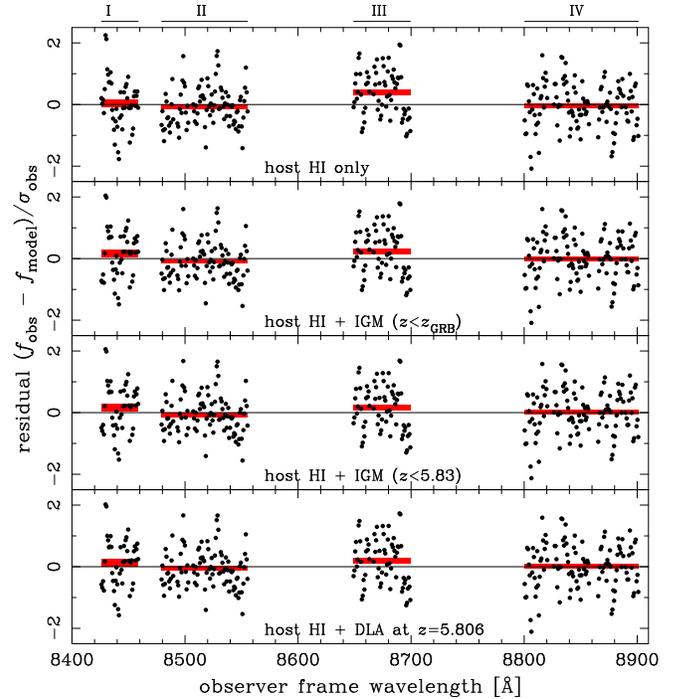}
\end{center}
\caption{The residuals of the data with respect to the model,
  normalized by the observational $1 \sigma$ error for each wavelength
  pixel, for the four models (host-only, IGM-$z_{\rm GRB}$, IGM-low$z_u$, and
  intervening DLA) whose model parameters are summarized in Table
  \ref{table:parameters}. The defined wavelength ranges I--IV are
  indicated at the top, and the red boxes indicate the mean residual
  within the four ranges, with the vertical width showing $\pm 1
  \sigma$ statistical uncertainty.  }
\label{fig:residual}
\end{figure}

We first examine the simplest model including only H\emissiontype{I}
in the host galaxy (the host-only model hereafter).  The best-fit
values of the three model parameters ($\beta$, $N_{\rm
  H\emissiontype{I}}^{\rm host}$, and $\sigma_v$) and the $\chi^2$
value are given in Table \ref{table:parameters}, and the fit residuals
$(f_{\rm obs} - f_{\rm model}) / \sigma_{\rm obs}$ of each pixel are
shown in the top panel of Fig. \ref{fig:residual}.  The 1$\sigma$
statistical errors for one model parameter in Table
\ref{table:parameters} were calculated by the standard procedure
(Press et al. 2007), i.e., finding the parameter range corresponding
to $\Delta \chi^2(p) \equiv \chi^2(p) - \chi^2_{\min} = 1$, where
$\chi^2_{\min}$ is the global minimum and $\Delta \chi^2(p)$ is the
minimum as a function of the parameter of interest, $p$, with the
other model parameters kept free (i.e., marginalized). For
convenience, we define the four wavelength ranges I--IV (indicated in
Figs. \ref{fig:spec_obs} and \ref{fig:residual}).

The most significant excess of the residuals of the host-only model is
found in the wavelength range III at a level of about 0.6\% of the
flux. The trends of increasing/decreasing residuals with increasing
wavelength in the ranges II/IV seem to be connected to the excess in
the range III, showing the overall systematic trend of concave
curvature spanning in the entire analysis wavelength range of
8400--8900 \AA.  We have examined various sources of systematic
uncertainties, but the 0.6\% level curvature in this narrow wavelength
range cannot be explained (see Section \ref{section:systematics}).

The $\chi^2$ value of this fit is 87.58, and statistically the chance
probability of getting this value is 2.7\% (i.e., 2.2$\sigma$
rejection) for the 68 data points and four model parameters (i.e., 64
degrees of freedom).  Here we examined this for a consistency check,
and later we will find that this model is disfavored by about
3$\sigma$ in comparison with other models. This sigma difference is
not surprising, because the total $\chi^2$ value and $\chi^2$
difference for different models are independent tests. For example, if
a model parameter is wrong in a model fit but the parameter is
affecting only a limited number of $\chi^2$ degrees of freedom, the
total $\chi^2$ value is insensitive and we need to see the $\chi^2$
difference when the model parameter is changed.

The concave curvature can be resolved by an absorption component that
becomes rapidly stronger at shorter wavelengths, and hence the redward
damping wing of another neutral hydrogen component on the sightline is
a good candidate to improve the fit.  Therefore we test the next
simplest model with diffuse IGM H\emissiontype{I} extending to the
same redshift as the GRB host, i.e., $z_{\rm IGM, u} = z_{\rm GRB}$
(the IGM-$z_{\rm GRB}$ model hereafter).  The best-fit of this model
with the new free parameter $f_{\rm H\emissiontype{I}}$ is presented
in Table \ref{table:parameters} and Fig. \ref{fig:residual}. The
best-fit is obtained at $f_{\rm H\emissiontype{I}} = 0.086$, and the
relative statistical significance against the host-only model can be
estimated again in the standard manner (Press et al. 2007) as follows.
Comparing the best-fit $\chi^2$ values of the host-only and
IGM-$z_{\rm GRB}$ models in Table \ref{table:parameters}, there is a
reduction of $\Delta \chi^2 = 8.92$ with one new parameter of $f_{\rm
  H\emissiontype{I}}$, where all the other parameters are
marginalized. Therefore the latter model is preferred at the
statistical significance of $8.92^{1/2} = 3.0\sigma$ from $\chi^2$
statistics with one degree of freedom.

\begin{figure}
\begin{center}
\includegraphics[width=120mm,angle=-90]{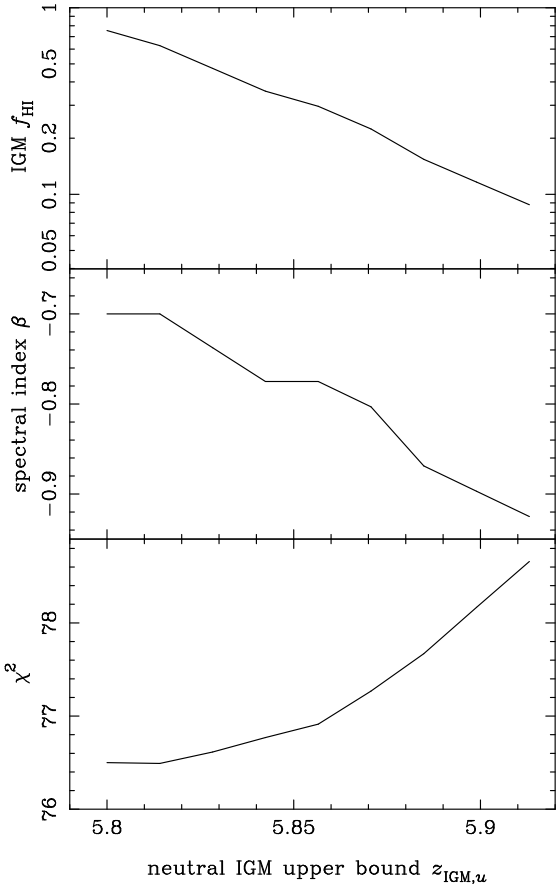}
\end{center}
\caption{The best-fit IGM neutral fraction $f_{\rm
    H\emissiontype{I}}$, power-law index $\beta$, and $\chi^2$ as a
  function of the upper bound of neutral IGM redshift distribution,
  $z_{\rm IGM,u}$. The lower bound is fixed at $z_{\rm IGM,l} =
  5.67$.}
\label{fig:free_z_u}
\end{figure}

However, the curvature of residuals is still evident for this model in
Fig. \ref{fig:residual}.  This may be further improved by allowing
lower values of $z_{\rm IGM,u}$, because such a model has a weaker
wavelength dependence of IGM absorption, and hence the effect can
reach to longer wavelength beyond the Ly$\alpha$ break, while keeping
the absorption around the break relatively weak.  Therefore we
performed a fit treating $z_{\rm IGM, u}$ as a further free model
parameter (the IGM-low$z_u$ model, hereafter), and the dependence of
the minimum $\chi^2$, best-fit $\beta$ and $f_{\rm H\emissiontype{I}}$
on $z_{\rm IGM, u}$ is shown in Fig. \ref{fig:free_z_u}, where the
other parameters of $N_{\rm H\emissiontype{I}}^{\rm host}$ and
$\sigma_v$ are marginalized by choosing values that minimize $\chi^2$
for each value of $z_{\rm IGM,u}$. We find a modest 
($\Delta \chi^2 \sim 2$, 1.4$\sigma$)
improvement of the fit by decreasing $z_{\rm IGM, u}$ down to $\sim
5.83$, with a higher value of $f_{\rm H\emissiontype{I}} = 0.47$.

It is interesting to note that this redshift is coincident with the
upper border of the region where the darkest GP troughs of Ly$\alpha$,
$\beta$, and $\gamma$ are found to this sightline ($z \sim$
5.67--5.83, Chornock et al. 2013), i.e., the region where we expect
the largest amount of IGM neutral hydrogen.  The physical proper
distance between $z = 5.83$ and the GRB host ($z = 5.913$) is 5.3
Mpc. The best-fit model parameters for $z_{\rm IGM, u} = 5.83$ are
presented in Table \ref{table:parameters}, and now the systematic
curvature has almost disappeared in the fit residuals
(Fig. \ref{fig:residual}).  This IGM-low$z_u$ model has two more
parameters of $f_{\rm H\emissiontype{I}}$ and $z_{\rm IGM, u}$ against
the host-only model, and $\Delta \chi^2 = 10.96$ is significant by the
chance probability of 0.0042 or $2.9 \sigma$ from the $\chi^2$
statistics of two degrees of freedom.  This result again indicates
that the host-only model is disfavored at $\sim 3 \sigma$ level.  On
the other hand, the difference between the two IGM models is not
statistically significant (1.4$\sigma$), and hence we cannot strongly
reject the IGM-$z_{\rm GRB}$ model.  The total $\chi^2$ value of the
model with $z_{\rm IGM, u} = 5.83$ is 76.62, corresponding to a modest
chance probability of 10.0\% for 62 degrees of freedom.

The best-fit host-only and IGM-low$z_u$ models are shown in
Fig. \ref{fig:spec_obs} compared with the observed flux.  To
understand these results qualitatively, we also show the flux
attenuation factor by various components of absorption in
Fig. \ref{fig:spec_model}.  The key point is that the IGM absorption
signature (especially with smaller $z_{\rm IGM,u}$) has a weaker
wavelength dependence than that by hydrogen in the host, which is
essential to resolve the curvature found in the residual spectrum of
the host-only fit.

Another possibility to improve the host-only fit is a DLA rather than
diffuse IGM. Therefore we test the intervening DLA model, in which the
DLA component at the redshift of the observed metal absorption system
($z = 5.806$) is added to the host-only model, but without including
diffuse IGM. The results are presented in Figs. \ref{fig:spec_model},
\ref{fig:residual}, and Table \ref{table:parameters}.  It is found
that this model gives a similarly good fit to the IGM-low$z_u$ model,
if the H\emissiontype{I} column density is $N_{\rm
  H\emissiontype{I}}^{\rm DLA} = 5.1 \times 10^{20} \ \rm cm^{-2}$,
which is about 10 times larger than $N_{\rm H\emissiontype{I}}$ in the
GRB host. We will further examine whether this intervening DLA model
is consistent with other spectral features in Section
\ref{section:DLA}.

\section{Examinations of Systematic Uncertainties}
\label{section:systematics}

\subsection{Systematics in the Spectrum Reduction}
The discrepancy between the host-only model and the observed data is
mainly because of the concave curvature of the data compared with the
best-fit model curve, extending in the entire wavelength range
analyzed (8400-8900 \AA) with an amplitude of 0.6\% of the continuum
flux (Fig. \ref{fig:residual}).  As mentioned above, the sensitivity
curve fit to the standard star is accurate by 0.1\% level, and the
0.6\% level difference of the spectral shape should be detectable.

To further test possible systematic uncertainties in the relative flux
calibration, we constructed another sensitivity curve by using another
CALSPEC standard star GD 153, which is one of the primary standards of
the CALSPEC system and was observed by Subaru one day before the
observation of GRB 130606A.  The airmass at the time of taking the
VPH900 spectrum of GRB 130606A is in between those for the two
standard stars (Feige 34 and GD 153).  The shapes of the two
sensitivity curves are consistent within 0.2\% accuracy, indicating
that the observed spectral shape is correct at 0.2\% level, taking
into account the variation of the standard star templates, airmass,
and atmospheric/instrumental conditions over the time scale of one
day.  The primary standards of the CALSPEC calibration system are the
model spectra of the three pure hydrogen white dwarfs including GD 153
(Bohlin 2007), and the model spectra have a smooth power-law profile
in 8400--8900 \AA, showing no complicated absorption structures.
Residuals of the HST data from the model spectra indicate that a
curvature of 0.6\% level in 8400--8900 {\AA} is highly unlikely to be a
result of the uncertainty about the CALSPEC template (see Fig. 3 of
Bohlin 2007).

In the wavelength range III, a broad enhancement of the spectral noise
is seen, which is a feature by the atmospheric O$_2$ emission
lines. If the spectrum is systematically biased by this effect, it
would have an important effect on our fitting result, because the
deviation of the data from the host-only model is the largest in this
range.  To check this we examined the fit residuals of the Subaru data
for the standard star with respect to the sensitivity curve fit, and
there is no systematic excess in this region. The noise level is
decreasing to larger wavelength within the range III, but the fit
residuals of the host-only model in Fig. \ref{fig:residual} do not
show any systematic wavelength dependence in this range.  The concave
curvature of the residuals ranging in the wide range of 8400--8900 \AA
\ cannot be resolved only by the O$_2$ feature in the range III.

We examined the atmospheric absorptions in 8500--8900 \AA \ using the
ATRAN database (http://atran.sofia.usra.edu/), and found that the
expected curvature produced by weak atmospheric absorption lines is
about 0.1\% in this wavelength range.  It should be noted that the
effect of atmospheric absorption should have been corrected if the
effect is the same for the standard star and the GRB
afterglow. Although a part of this effect may remain uncorrected by a
possible systematic change of atmospheric conditions at the observing
times for the star and the afterglow, it is impossible to explain the
0.6\% level curvature of the host-only fit residuals.

We also examined the effect of slit/aperture loss of afterglow light.
The slit loss of light changes the relative flux of the spectrum by
the atmospheric dispersion effect, but we confirmed that this is less
than 0.05\% within the wavelength range analyzed.  We found that the
image size along the spatial direction changes with wavelength, which
is likely due to wavelength dependent seeing and/or chromatic
aberration.  Therefore we quantitatively examined the systematic
change of the spectrum by aperture loss. (The aperture size for the
GRB spectra is smaller than that for the standard star.)  We found
that this changes the relative flux at 8400 and 8900 \AA \ by $\sim$
2.6\%, but the wavelength dependence is monotonic and well described
by a power-law, having an effect of changing $\beta$ by 0.45. The
analyzed spectrum has been corrected about this before the model
fittings, and hence correction for this effect is not necessary for
the $\beta$ values presented in this paper.  Since this effect is well
described by a power-law, it is unlikely to be the origin of the
residuals showing a curvature.

\subsection{Systematics in Modelings and Fitting Procedures}
We have removed the wavelength regions showing visually discernible
absorption features from our analysis, but weaker absorption lines may
affect the fit results.  Since it is difficult to test the effect of
weak lines that cannot be recognized, we instead performed the
fittings including the four relatively weak absorption features at
8492, 8505, 8824, and 8889 \AA, which were removed in our baseline
analysis.  Some of these may be statistical fluctuation rather than
real lines. The best-fit $\chi^2$ values for the host-only and
IGM-$z_{\rm GRB}$ models are now 134.69 and 123.11 for 69 data points, and
the preference for the IGM model ($\Delta \chi^2 = 11.58$) is not
significantly changed.  We also tested possible systematic effects
induced by large noise in the regions of atmospheric airglow lines.
We repeated the fittings after removing the narrow airglow regions
centered at 8435, 8455, 8495, 8505, 8542, 8552, 8830, 8840, 8852,
8870, and 8888 \AA.  (These regions can be seen as bumps in the error
curve shown in Fig. \ref{fig:spec_obs}.)  The $\chi^2$ values of the
two models are 60.17 and 51.73 for 45 data points, and $\Delta
\chi^2 = 8.44$ is still significant at 2.9$\sigma$.

We used the central redshift of $z_{\rm GRB} = 5.9131$ for the
H\emissiontype{I} gas in the GRB host, based on the redshifts of the
metal absorption lines measured in the same spectrum.  The
C\emissiontype{II} line shows a relatively low redshift of $z =
5.9127$, and the uncertainty about the host redshift can be tested by
adopting this value for $z_{\rm GRB}$.  We then found the best-fit
$\chi^2$ values of 86.13 and 79.30 for the host-only and IGM-$z_{\rm
  GRB}$ models, respectively, and $\Delta \chi^2 = 6.83$ is still
significant at 2.6$\sigma$.

We considered only a single power-law for the intrinsic afterglow
spectrum before absorption effects, though afterglows sometimes show
concave breaks in their spectral energy distribution (SED), which may
explain the curvature found in the residuals of the host-only
fit. However, if a spectral break is close to the observed wavelength
range, we expect from the afterglow theory a significant time
evolution of SED by the passage of the break in an observing band
(Zhang 2007). As shown in Fig. \ref{fig:NIR_SED}, the reported
near-infrared colors are fairly constant from 36 min to 35 hr after
the burst, and no evidence for spectral breaks other than the
Ly$\alpha$ break is seen.  If the 0.6\% curvature in 8400--8900 \AA
\ is represented by a quadratic term in the $\lg f_\lambda$ versus
$\lg \lambda$ relation, it would result in a change of $\Delta \beta =
1$ by just a 7\% change of wavelength $\lambda$. Such a strong break
is not expected from the afterglow theory (Zhang 2007), and
inconsistent with the agreement of $\beta$ for our fits and
near-infrared colors.

Since GRB afterglows are time-variable, we also examined the time
variability of the reduced spectrum, by reproducing two spectra using
the first and second halves of the integrations (1--4 and 5--9 of
Table \ref{table:obs}).  We confirmed that there is no statistically
significant change in the spectra, and the concave curvature of
the residuals for the host-only model is seen in both the spectra.

\subsection{Comparison with A Previous Study}
Chornock et al. (2013) also reported an analysis of the damping wing
shape of the same GRB afterglow taken by another telescope, and they
found no statistically significant preference for the diffuse IGM
component.  This result is apparently different from ours, although
their upper limit ($f_{\rm H\emissiontype{I}} < 0.11$ at $2\sigma$) is
consistent with our best-fit of $f_{\rm H\emissiontype{I}} = 0.086$
for the IGM-$z_{\rm GRB}$ model.  The origin of this difference is
difficult to identify, but a clear difference of their analysis from
ours is the index of unabsorbed power-law spectrum: $f_\lambda \propto
\lambda^{-0.01}$, i.e., $\beta = -1.99$. Such a small value is
significantly different from those measured by near-infrared colors,
and it is not supported either from our $\beta$-free fitting analysis,
giving an unacceptably large $\chi^2$ of $\sim 470$.  The fit by
Chornock et al. indeed shows a concave curvature of the observed
spectrum compared with the fitted model (see their Fig. 2), and this
curvature and their small (red) $\beta$ value could be a result of a
power-law fit in a wavelength region where the IGM/DLA component is
not negligible.

\section{Examination of the Intervening DLA Scenario}
\label{section:DLA}

As already presented, the evidence for the intervening
H\emissiontype{I} absorption presented above can be explained by a DLA
associated with the observed metal absorption system at $z = 5.806$,
if the H\emissiontype{I} column density is $\lg N_{\rm
  H\emissiontype{I}}^{\rm DLA} = 20.704^{+0.066}_{-0.067}$.  Here we
examine whether this scenario is consistent with other observed
features of the GRB 130606A afterglow spectrum.

\subsection{Expected Number of DLAs Along the Sightline}
The expected number of such a DLA along the line of sight can be
estimated as follows. A modest extrapolation of the DLA statistics up
to $z \sim 5$ (Wolfe et al. 2005; Songaila \& Cowie 2010) suggests the
DLA number distribution per unit redshift $dN_{\rm DLA}/dz \sim 0.72$
at $z = 6$ for $\lg N_{\rm H\emissiontype{I}} \ge 20.3$.  Assuming the
column density distribution as $N_{\rm DLA}(> N_{\rm
  H\emissiontype{I}}) \propto N_{\rm H\emissiontype{I}}^{-0.85}$ again
from the lower-$z$ sample, the probability of finding such a DLA in
the redshift range between 5.806 and 5.913 is estimated to be $\sim
0.03$.  Therefore the chance probability of finding such a DLA is
rather small, though it is not negligible.

\subsection{Lyman Series Absorption Features}
\begin{figure*}
\begin{center}
\includegraphics[width=120mm,angle=-90]{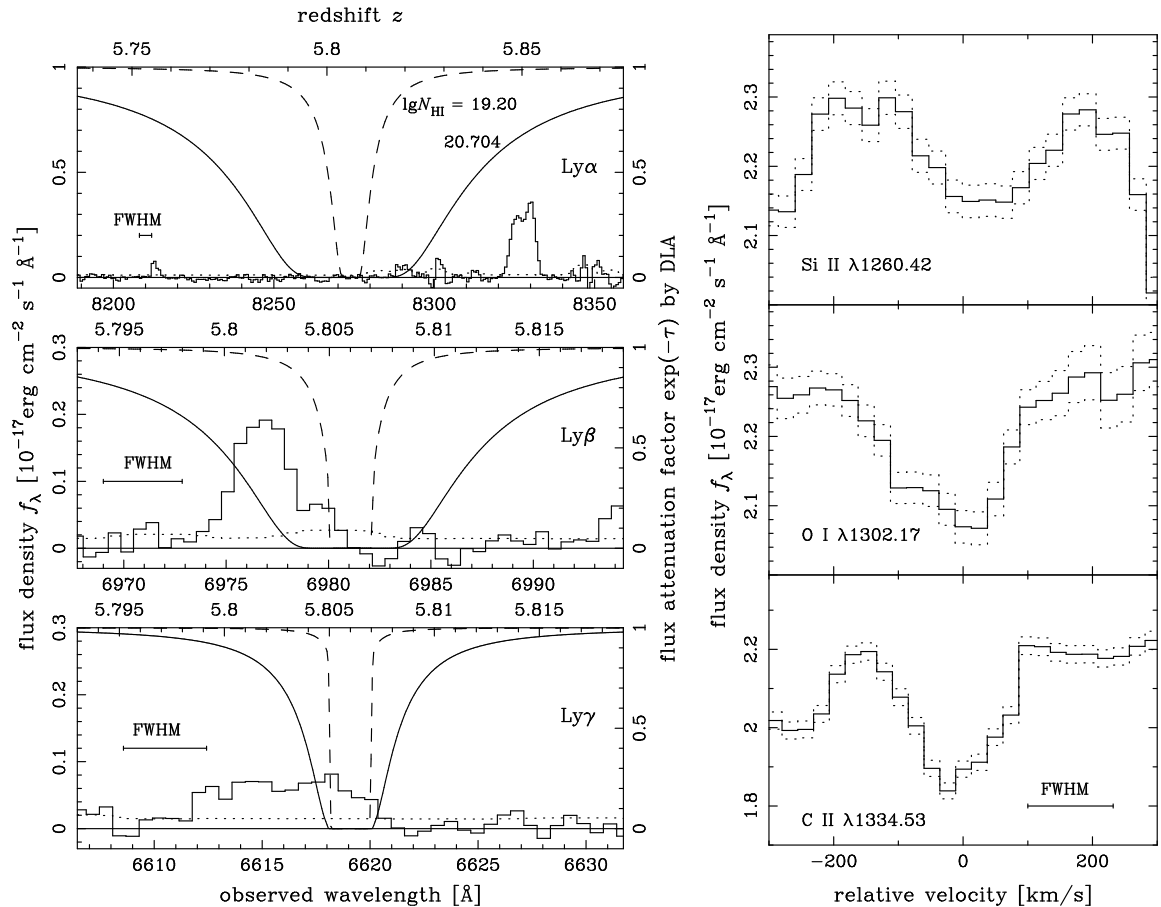}
\end{center}
\caption{Left panels: the light transmission profiles around Lyman
  $\alpha/\beta/\gamma$ at $z = 5.806$.  The solid and dotted
  histograms are the observed spectrum and its 1$\sigma$ noise (VPH900
  for Ly$\alpha$ and VPH650 for Ly$\beta$/$\gamma$).  The solid and
  dashed curves are flux attenuation factor $e^{-\tau}$ of Lyman
  series absorption (see the right-hand-side ordinate for scales) by a
  DLA located at $z = 5.806$, with two different H\emissiontype{I}
  \ column densities (in units of cm$^{-2}$) as indicated in the top
  panel.  Right panels: the metal line profiles of the $z = 5.806$
  absorption system, shown as a function of relative velocities.  The
  dotted histograms are showing $\pm 1 \sigma$ errors.  The spectral
  resolutions (FWHM) are also indicated by the horizontal bars.  }
\label{fig:Ly_prof}
\end{figure*}

The existence of such a large column DLA at $z = 5.806$ can be further
constrained by the spectral profiles around Lyman series absorption at
this redshift, because absorption by H\emissiontype{I} in the DLA
would erase the transmission of light around this redshift.  The
spectrum around Ly$\alpha$, $\beta$, and $\gamma$ are shown in
Fig. \ref{fig:Ly_prof}, with the expected absorption factor by the DLA
having $\lg N_{\rm H\emissiontype{I}}^{\rm DLA} = 20.704$. Here, the
velocity distribution of H\emissiontype{I} in the DLA is modeled as
follows.  The profiles of the three metal absorption lines
(Si\emissiontype{II}, O\emissiontype{I}, C\emissiontype{II}) at this
redshift are also shown in Fig. \ref{fig:Ly_prof}, and their widths
are consistent with the spectral resolution (3.9 \AA \ FWHM),
indicating narrow intrinsic widths.  The relative velocity difference
of these lines are at most $\sim 50$ km/s.  The metal absorption line
profile of typical DLAs is a mixture of multiple components with a
velocity width of $\sigma_v \sim 4$--7 km/s for one component, and the
relative velocity width $\Delta v$ of different components is widely
distributed in $\sim$20--200 km/s (Wolfe et al. 2005;
Dessauges-Zavadsky et al. 2003).  Therefore, as a reasonable model, we
adopt $\sigma_v = 5.5$ km/s for each single component and assumed that
these components are uniformly distributed in the relative velocity
range of $\Delta v = -20$ to +20 km/s with respect to $z = 5.806$.

The light transmission around the Lyman series in
Fig. \ref{fig:Ly_prof} must be consistent with the predicted
attenuation factor by the DLA, and the transmissions around Ly$\alpha$
and $\gamma$ satisfy this requirement, if the spectral resolution is
taken into account.  However, a large transmission flux centered at
6976.97 \AA \ is found near Ly$\beta$ with an integrated flux of $(1.6
\pm 0.1) \times 10^{-17} \ \rm erg \ cm^{-2} s^{-1}$.  This
transmission profile has a width comparable with the spectral
resolution, indicating a narrow intrinsic width. Furthermore, the
light attenuation by the assumed DLA becomes rapidly stronger beyond
6977 \AA, which also requires that the transmission is coming from a
narrow ($\lesssim$ 1 \AA) wavelength window. This means that the
transmission flux density must be $f_{\lambda} \ge 1.6 \times 10^{-17}
\ \rm erg \ cm^{-2} s^{-1}$ \AA$^{-1}$, which should be compared with
the unabsorbed continuum flux $3.6 \times 10^{-17} \ \rm erg \ cm^{-2}
s^{-1}$ \AA$^{-1}$ estimated by the continuum flux at 9800 \AA \ and
an extrapolation by using the best-fit power-law $\beta = -0.81$ of
the intervening DLA model (Table \ref{table:parameters}).  Here, these
flux estimates of the transmission and unabsorbed continuum were made
from the 300R spectrum that covers Ly$\beta$ and 9800 \AA
\ simultaneously, and hence we do not have to consider the time
evolution of the flux.

This result means that the IGM transmission fraction must be larger
than 1.6/3.6 = 0.44 at 6977 \AA, but this is inconsistent with the
transmission allowed by Ly$\beta$ absorption of the DLA that we are
now considering, 0.18$\pm$0.05.  Hence the particular case that the
DLA at $z = 5.806$ is responsible for the unexplained residuals of the
host-only model is excluded.  We examined possible systematic
uncertainties about the relative flux calibration at 6977 and 9800 \AA
\ (by aperture loss, atmospheric dispersion, guide error, standard
star, and wavelength-dependent seeing), and it should be less than
10\%.  A change of $\beta$ by 0.1 results in just a 3\% change of the
unabsorbed flux at the Ly$\beta$ wavelength.  It should also be noted
that the modeling about velocity distribution of H\emissiontype{I} in
the DLA hardly affects this argument, because the 6977 \AA
\ transmission is far from the resonant Ly$\beta$ wavelength and the
attenuation is simply determined by the damping wing.

Therefore, the H\emissiontype{I} column density of the $z = 5.806$
system must be lower than the value required to explain the residual
of the host-only model.  If we assume that the $z = 5.806$ system has
a similar metallicity to the GRB host galaxy, a reasonable amount of
the H\emissiontype{I} column density would be $\lg N_{\rm
  H\emissiontype{I}} \sim 19.2$, since the metal-line equivalent
widths (Si\emissiontype{II}, O\emissiontype{I}, and
C\emissiontype{II}) of the $z = 5.806$ system are about 3--9 times
lower than those found in the GRB host galaxy (Chornock et al. 2013).
Flux attenuation factors assuming this H\emissiontype{I} column
density are also plotted in Fig. \ref{fig:Ly_prof}, and in this case
Ly$\beta$ absorption at 6977 \AA \ is negligible, thus allowing the
existence of the observed transmission there. Such a small
H\emissiontype{I} column density does not affect the fitting of the
Ly$\alpha$ damping wing at the GRB redshift.

Finally, we note that the simple diffuse IGM models with uniform
$f_{\rm H\emissiontype{I}} \gtrsim 0.1$ is also mathematically excluded
by the light transmission at 6977 \AA, because of the large predicted
Ly$\beta$ GP optical depth of $\tau_{\beta} = 7.2 \times 10^4 f_{\rm
  H\emissiontype{I}}$ at $z = 5.802$ using the absorption oscillator
strength ratio of $f_{\beta}/f_{\alpha} = 0.190$. However, in reality
the neutral fraction cannot be perfectly homogeneous, and many
transmission spikes are expected at locally ionized bubbles along
the sightlines, as seen in the spectrum of this GRB (Chornock et
al. 2013). Therefore, a transmission spike at 6977 {\AA} does not
exclude any realistic models of diffuse IGM absorption.

\subsection{Metallicities}
Finally, even if the large column of $\lg N_{\rm
  H\emissiontype{I}}^{\rm DLA} = 20.704$ is associated with the $z =
5.806$ system, the metallicity of this system must then be extremely
low.  From the reported equivalent widths of the Si\emissiontype{II},
O\emissiontype{I}, and C\emissiontype{II} lines (Chornock et al. 2013)
and this H\emissiontype{I} column, we estimate the metallicity of
[Si/H], [O/H], and [C/H] to be $-3.68 \pm 0.10$, $-3.27 \pm 0.10$, and
$-3.51 \pm 0.12$, respectively, in the optically-thin limit.  Here we
followed the usual practice for DLA studies and did not apply
ionization correction, because these low-ionization atoms are the
dominant phase of these elements in interstellar medium.  These lines
may be saturated, and hence we estimated the metal column density
considering the saturation effect, assuming that these lines have a
single component of $\sigma_v = 5.5$ km/s.  Then the metallicities are
increased to $-3.48^{+0.15}_{-0.16}$, $-2.69^{+0.44}_{-0.31}$, and
$-3.19^{+0.30}_{-0.26}$, respectively.  (Note that the saturation
effect would be weaker than this, if these lines are composed of
multiple components.)  

The saturation effect is not large for the weak line of
Si\emissiontype{II}, and [Si/H] of this system must be significantly
lower than the lowest value ($\sim -2.7$) found for DLAs at lower
redshifts up to $z \sim 5$ (Rafelski et al. 2012). In comparison with
a theoretical model of Finlator et al. (2013), the
saturation-corrected column density of $\lg N_{\rm O\emissiontype{I}}
= 14.7$ for a DLA having $\lg N_{\rm H\emissiontype{I}}^{\rm DLA} =
20.704$ at $z \sim 6$ is outside the $N_{\rm
  O\emissiontype{I}}$-$N_{\rm H\emissiontype{I}}$ distribution of 99\%
sightlines. We have already argued that the Ly$\beta$ profile excludes
the particular case of the DLA at $z = 5.806$ as an explanation of the
residuals found in the host-only fit, and one may consider the
possibility of a DLA at a different redshift to save the intervening
DLA scenario. However, in this case the DLA metallicity must be
further lower than the values estimated above, because there are no
detected metal absorption lines.

\section{Discussion and Conclusions}
\label{section:conclusions}

We presented an analysis of the redward Ly$\alpha$ damping wing found
in the optical afterglow spectrum of GRB 130606A at $z = 5.913$, for
the purpose of getting constraints on cosmic reionization.  We used
the spectrum in the narrow wavelength range of 8400--8900 \AA, which
was specially reduced for the particular purpose of the damping wing
analysis to minimize systematic uncertainties. The minimal model
including the unabsorbed baseline power-law and H\emissiontype{I}
absorption in the host galaxy does not give a good fit, leaving
residuals showing a concave curvature whose amplitude is about 0.6\%
of the flux.  Such a curvature in the short wavelength range cannot be
explained by extinction at the host galaxy, if we adopt the standard
extinction curves, because the standard curves have monotonic
wavelength dependence in this wavelength range, and change of
extinction is almost perfectly absorbed by according change of the
power-law index $\beta$. We cannot exclude an unknown anomalous
behavior of extinction curve in this wavelength range, but if this is
the case, it would have interesting implications for the physics of
interstellar dust grains in the early universe. We have also examined
various sources of systematic uncertainties in the observed spectrum
and model fitting procedures (Section \ref{section:systematics}), but
we cannot resolve the residuals of the host-only model.

This motivated us to test intervening H\emissiontype{I} absorption
components along the line of sight, and we found that both diffuse IGM
and a DLA at the redshift of an observed metal absorption system ($z =
5.806$) can reduce the residual, with statistical significances of
about 3$\sigma$.  This is the first evidence for intervening neutral
hydrogen outside the host galaxy, found in the red damping wing of GRB
afterglow spectra.

We then examined the consistency of the DLA scenario with other
observed features in the spectrum (Section \ref{section:DLA}).
Probability of finding such a DLA along the sightline is rather small
($\sim$3\%).  The particular case of the $z = 5.806$ DLA can be
excluded from the light transmission profile around Ly$\beta$.  The
metallicity argument also severely constrains this scenario.  Even if
the DLA is associated with the $z = 5.806$ metal absorption system,
its silicon abundance must be about 1/2000 times the solar level,
which is significantly lower than the lowest known metallicity ($\sim
1/500$) of DLAs at $z \lesssim 5$ and theoretical expectation at $z
\sim 6$.  If the DLA is located at a redshift different from 5.806, no
detected metal lines require an even lower metallicity.  Therefore we
conclude that the intervening DLA scenario does not provide a
reasonable explanation for the residuals of the host-only model. If
the DLA scenario was correct, it would imply a rapid increase of the
incidence of extremely low metallicity DLAs around $z \sim 6$, which
would give an interesting constraint on the theory of early galaxy
formation.

Hence, there are good reasons to seriously consider the possibility
that the observed absorption feature is made by neutral hydrogen in
diffuse IGM.  If correct, it can be interpreted as the remnant of
neutral IGM at the relatively late phase of reionization.  The
best-fit IGM neutral fraction varies depending on the upper redshift
bound, from $f_{\rm H\emissiontype{I}} \sim 0.1$ for $z_{\rm IGM, u} =
z_{\rm GRB}$ to $\sim 0.5$ for $z_{\rm IGM, u} = 5.83$ (the upper
bound of the darkest GP troughs to this sightline).  These values are
interestingly close to unity, but quantitatively should be taken with
care, because the adopted IGM model (uniform density and $f_{\rm
  H\emissiontype{I}}$) is obviously too simple (McQuinn et al. 2008;
Mesinger \& Furlanetto 2008).  Ly$\alpha/\beta/\gamma$ transmission
spikes are found between the dark GP troughs from $z = 5.67$ to the
GRB redshift (Chornock et al. 2013), and $f_{\rm H\emissiontype{I}}$
cannot be such a high value at the spike regions. On the other hand,
theory predicts that reionization proceeds inhomogeneously (e.g.,
Iliev et al. 2006), and high $f_{\rm H\emissiontype{I}}$ gas can be
hidden in clumpy IGM corresponding to dark GP troughs.  A comparison
with more realistic theoretical models would place more quantitative
constraints on the reionization history.

It should be noted that the red damping wing signature of absorption
by intervening neutral hydrogen has also been reported for a quasar at
an even higher redshift of $z = 7.085$ (Mortlock et al. 2011; Bolton
et al. 2011), whose column density is similar to that found for GRB
130606A.  Indirect evidence for the damping wing by highly neutral IGM
has also been claimed in quasar spectra at $z \sim 6$ (Schroeder et
al. 2013).  Since GRBs are less biased tracers and do not alter the
surrounding ionization status, it is reasonable to expect a higher
$f_{\rm H\emissiontype{I}}$ in regions around GRB host galaxies than
quasars at the same redshift.  These quasar results and the new result
of GRB 130606A suggest that the cosmic reionization is not yet
complete at $z \sim 6$, and we are now starting to detect the remnant
neutral hydrogen from the universe before reionization in various
environments.

Our result demonstrates that damping wing analyses of
GRB afterglows give useful constraints on reionization. It is
sensitive to the IGM neutral fraction of $f_{\rm H\emissiontype{I}}
\sim 0.1$ for H\emissiontype{I} gas close to the host, or $\sim 0.5$
for that separated from the host by $\Delta z \sim 0.08$ or 5 Mpc in
proper distance.  More high quality spectra of GRB afterglows in the
reionization era would give us a more complete picture of reionization
history in the near future, which will be made possible by planned
30--40m class telescopes.

\bigskip
The spectrum used in this work was collected at the Subaru Telescope,
which is operated by the National Astronomical Observatory of Japan.
This work, as well as the upgrade of Subaru/FOCAS in 2010, was
financially supported by the MEXT Grant-in-Aid for Scientific Research
on Priority Areas (19047003).

%%%
% See the manual for the detail.
%%%

\end{document}